\documentclass[aps,pra,amssymb,amsmath,superscriptaddress]{revtex4}
\usepackage{graphicx,color,braket}
\usepackage{amsmath}
\usepackage{amssymb}
\newcommand{\bla}{\color{black}}

\usepackage[colorlinks = true]{hyperref}
\usepackage[T1]{fontenc}

\begin{document}
	
\title{Ping-pong quantum key distribution with trusted noise: non-Markovian advantage}  \author{
  Shrikant Utagi}  
\email{shrik@poornaprajna.org}
 \affiliation{Poornaprajna    Institute   of   Scientific
  Research, Bangalore -  562164, India} 
\affiliation{Graduate  Studies,  Manipal Academy of Higher Education,  Manipal-576104}
\author{R.     Srikanth}  
\email{srik@poornaprajna.org}
 \affiliation{Poornaprajna    Institute   of
  Scientific     Research,     Bangalore      -     562164,     India}
 \author{Subhashish       Banerjee}
\affiliation{Indian Institute of Technology,  Jodhpur -   342037,  India.}
\email{subhashish@iitj.ac.in}
\begin{abstract}
	The ping-pong protocol adapted for quantum key distribution is studied in the trusted quantum noise scenario, wherein the legitimate parties can add noise locally. For a well-studied attack model, we show how non-unital quantum non-Markovianity of the added noise can improve the key rate.  We also point out that this noise-induced advantage cannot be obtained by Alice and Bob by adding local classical noise to their post-measurement data.
\end{abstract}

\maketitle

\section{Introduction}
Quantum key distribution (QKD) protocols are known to offer information theoretic security  of information, unlike their classical counterparts which can only offer computational security. Over the time, a number of QKD  protocols have been proposed (cf. the review \cite{shenoy2017quantum}), since their foundation was laid over three decades ago by Bennett and Brassard \cite{bennett1984quantum}. Whilst QKD protocols typically involve the probabilistic generation of a secret key, \cite{bostrom2002deterministic} proposed a deterministic version thereof using entanglement in a two-way protocol (called the ``Pingpong protocol'', described below), but it turns out that the idea can also be realized without entanglement \cite{lucamarini2005secure}. Certain attacks or modifications to the Pingpong protocol were proposed in \cite{wojcik2003eavesdropping,cai2004improving,cai2006,zhang2004improving}, which were analyzed in \cite{bostrom2008security}. Subsequently, further modifications or attacks on the Pingpong protocol were studied by other authors \cite{vasiliu2011non,zawadzki2012security,zawadzki2012improving,li2012improved,han2014security}

Noise is especially detrimental to quantum information processing, given the fragility of quantum resources \cite{banerjee2018open}. Yet, recently, there have been a few reports pointing out that the addition of classical or quantum noise by information sender Alice or receiver Bob can be advantageous to QKD \cite{renner2005information,bruss2008optimal,mertz2013qkd}. Here, we shall refer to such user added noise as ``trusted''. Note that this terminology differs from that used by \cite{usenko2016trusted}, who in the context of continuous variable QKD protocols \cite{patron2009continuous} refer to noise that is security breaking as ``untrusted'' and noise that is merely key rate reducing as ``trusted''.

Quantum non-Markovianity of noise is the quantum analogue of classical memory effects and manifests itself through the backflow of quantum information or increase in the distinguishability of two states subjected to a noisy channel \cite{RHP14,kumar2017nonmarkovian,li2018concepts}, though we may reasonably posit weaker manifestations of quantum non-Markovianity (cf. \cite{shrikant2020temporal,pollock2018operational}). Thus, it is intuitive to expect that quantum non-Markovianity can be helpful to information processing \cite{sharma2018decoherence,thapliyal2017quantum}, especially at low temperatures \cite{weiss2008quantum,vega2017dynamics} . However, this is by no means automatic (cf. e.g., \cite{rossi2018non}).


In an earlier work it was shown \cite{sharma2018decoherence} that non-unital noise helps cryptographic security for QKD based on the Pingpong communication protocol for a specific attack, essentially because the noise turns out to be more detrimental for Eve than Alice and Bob. In this paper, we show that non-Markovianity can further boost the advantage given by the non-unitality of quantum channels under certain circumstances. As before,  unital channels provide no advantage.  We consider two different scenarios in which amplitude damping noise is deliberately applied by a legitimate party (Bob, specifically) before a Bell measurement, and study the increase in secure key rate. In both cases, we find that if the quantum noise is non-Markovian, then the secure key rate increases significantly in comparison to Markovian noise in certain time ranges. 

There do not seem many works that have explored this practically useful aspect. Notably, Ref. \cite{mertz2013qkd} shows that deliberately adding depolarizing noise increases secure key rate for BB84 \cite{bennett1984quantum} and for entanglement based six-state protocols \cite{bruss1998optimal,bruss2000quantum}. This was somewhat inspired from the work \cite{bruss2008optimal} where for the six-state protocol, white noise added by the sender to the message qubit either prior to sending the qubit or prior to measurement on the qubit, gives rise to an increased secure key rate in the sense we consider in this paper.

This paper is arranged as follows. In Section \ref{sec:2}, we introduce the protocol, which is the ``ping-pong'' communication protocol adapted for QKD. In Section \ref{sec:3} we discuss the phenomenologically motivated model of amplitude damping noise and describe how it can be added during the protocol. We consider in Section \ref{sec:3a}  the first scenario involving a single-qubit noise, and in Section \ref{sec:3b}, the second scenario involving two-qubit incoherent noise. In section \ref{sec:proof} we show that the noisy joint statistics cannot be simulated by locally adding classical randomness to the noiseless joint quantum statistics of the protocol. Then, we conclude in Section \ref{sec:conclu}. 

\section{The basic protocol and the optimal individual attack \label{sec:2}}

The Pingpong protocol, adapted as a scheme for QKD, runs as follows:  Bob prepares the Bell state $\ket{\psi^{+}} = \frac{1}{\sqrt{2}}(\ket{01} + \ket{10})$, in particular pair of photons entangled in the polarization degree of freedom, out of which he sends one photon (travel photon) to Alice through a quantum channel, ideally assumed to be noiseless and lossless.  Alice then encodes the travel qubit by applying either $I$ or $\sigma_z$ with probability $\frac{1}{2}$, and sends it back to Bob. Once the travel qubit returns to Bob, he is left with either of the two Bell states $\ket{\psi^{\pm}} = \frac{1}{\sqrt{2}}(\ket{01}\pm \ket{10})$, corresponding to the bit 0 or 1 encoded by Alice, which he distinguishes through a Bell measurement. 

In the original ping-pong quantum direct communication protocol, the security requires alternating between the above message mode and a control mode, wherein Alice measures the travel qubit for error checking, and does not return it. Here,  for the requirement of QKD, we drop the control mode and consider only the message mode.  As a security check, both parties compute the quantum bit error rate (QBER) by sampling a fraction of the qubits transmitted. On them, Alice announces her encoded bit and Bob announces the Bell state he detected.  The fraction of cases where their records differ is an estimate of QBER, and a potential indicator of eavesdropper Eve's presence. If QBER is found to be less than a threshold value, they proceed ahead with key distillation, or else they abort the protocol.
\begin{figure}
	\centering
	\includegraphics[width=0.8\textwidth]{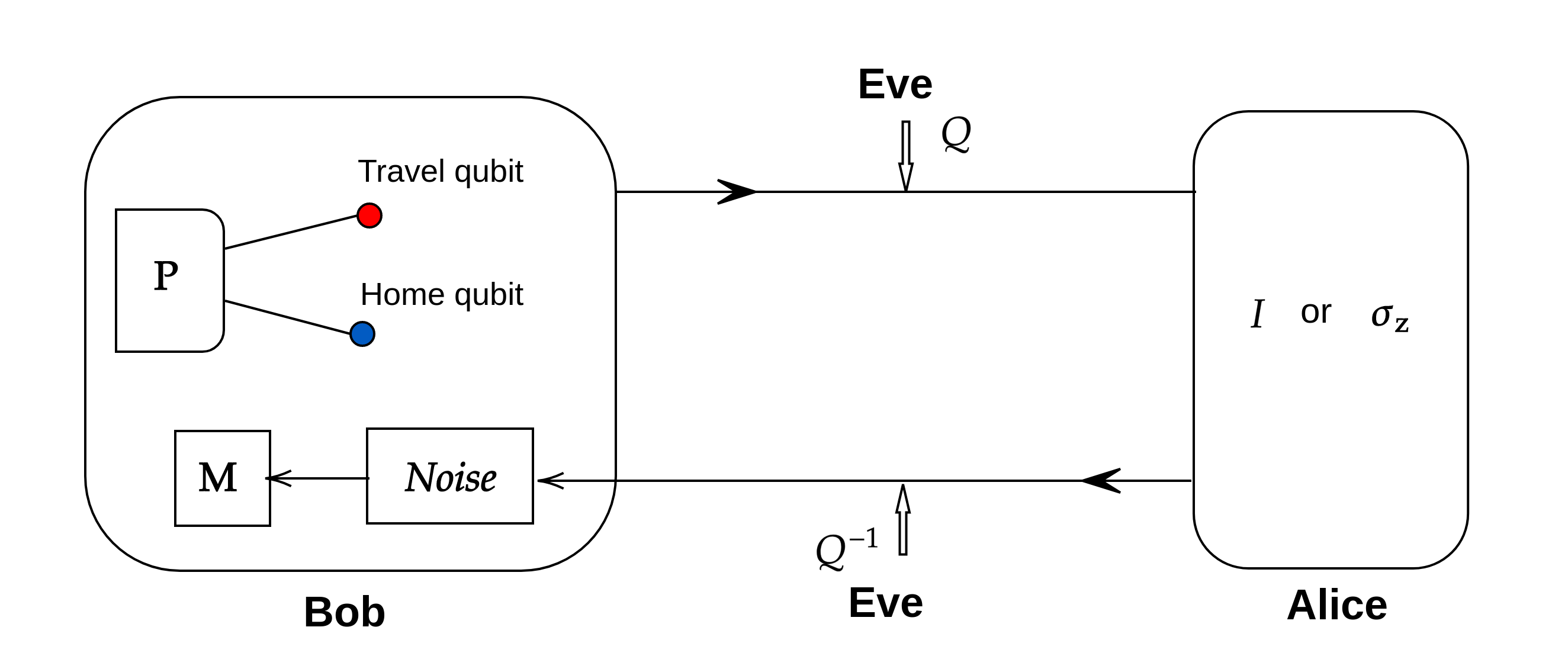}
	\caption{(Color online.) General scheme of the Pingpong protocol \cite{bostrom2002deterministic}: Bob prepares an entangled qubit pair in polarization degrees of freedom, and transmits the travel qubit to Alice and retains the home qubit. All channel noise is conservatively assumed to be due to Eve's attack. In addition, Bob adds noise to the qubit(s) prior to his measurements.  The same layout is used when the protocol is adapted for QKD, except that the control mode is dropped (cf. text) } 
	\label{fig:fig1}
\end{figure}

The interesting aspect of the Pingpong protocol is that in the ideal case, Eve only finds the onward and return photons to be in the maximally mixed state. Wojcik \cite{wojcik2003eavesdropping} proposed a strategy by attacking the onward and return legs.  In this attack, Eve includes two ancillary particles, the first (labelled $x$) prepared in a vaccum state, denoted $\ket{2}$, and the other (labelled $y$) in the state denoted $\ket{y} = \ket{0}$. Then the composite initial state is $\ket{\Psi}^{\rm initial}_{\rm htxy} = \ket{\psi^+}\ket{2}_x \ket{0}_y$, where $h$ and $t$ are labels for ``home'' and ``travel'' qubit states respectively. \bla In the onward leg, Eve  attacks the travel qubit by applying the
operation given by:
\begin{equation}
Q_{txy}: \left. \begin{array}{c}
\ket{020} \\
\ket{021} \\ \ket{120} \\ \ket{121}
\end{array} \right\}
\longrightarrow
\left\{ \begin{array}{c}
\ket{002}+\ket{201} \\
\ket{002}-\ket{201} \\ \ket{210}+\ket{112} \\ \ket{210}-\ket{112}
\end{array} \right.
\end{equation}
with  CPBS denoting
the ``controlled polarization beam splitter'' operation. On the return leg (after Alice's encoding action), Eve  applies the operation $Q_{txy}^{-1}$  on the travel qubit and forwards it to Bob.

After the end of the quantum round, Bob receives the final states 
$
\ket{\Psi}_{\rm fin} = \frac{1}{\sqrt{2}} (\ket{012j}+\ket{1020}),
$
with $j \in \{0,1\}$, corresponding to Alice's operation $\hat{O}_j \in \{I, Z\}$. The joint probabilities of Alice, Eve and Bob, $P_{\rm AEB}$, are found to be
\begin{align}
P_{000}=\frac{1}{2}; \quad
P_{1jk}=\frac{1}{8},
\label{eq:noiselessjp}
\end{align}
for $j, k \in \{0,1\}$.

The secure (or secret) key rate for this individual attack on each travel by Eve is lower bounded by $k_{\min} = I(A:B) - \chi(A:E)$, where $I(A:B)$ is the mutual information between the trusted parties Alice and Bob, and $\chi(A:E)$ is the Holevo information between trusted party Alice and malicious Eve. In practice, the key rate may be as high as determined  $k_{\max} = I(A:B) - I(A:E)$.  For the noiseless case of (\ref{eq:noiselessjp}) , it turns out that $I(A:B) = I(A:E) = \chi(A:E) \approx 0.31$ implying that the key rate vanishes and that Eve's attack strategy is indeed optimal for this protocol. 

\section{Noise advantage \label{sec:3}}

In general, it is known that noise can degrade the quantum information processing tasks, in particular QKD. 
In Ref. \cite{sharma2018decoherence}, we pointed out the surprising fact of advantage that noise can bestow on QKD. Here we extend that analysis, by including the role of memory in the quantum dynamics. Because the noise brings an advantage, we can visualize the scenario wherein Bob (or Alice) deliberately adds such beneficial noise to the particles.

We consider two scenarios, wherein Bob, before making Bell measurements on the entangled pair of particles, but after receiving the travel qubit, introduces noise into the system. In the first case, he subjects the travel qubit alone to an optical setup that simulates AD. In the second case, he subjects both the photons to noisy devices in the above manner.
In both scenarios, Eve is still assumed to act according to the attack described in Section \ref{sec:2}. Note that we may also assume that the noise occurs naturally because of Bob's noisy devices, and he merely takes advantage of it.

For the noisy dynamics introduced by Bob, we consider a non-Markovian amplitude damping (NMAD) channel, modeled by damped Jaynes-Cummings model with operator-sum representation given by the Kraus operators \cite{srikanth2008squeezed}
\begin{equation}
E_{0}^{A}=\left[\begin{array}{cc}
1 & 0\\
0 & \sqrt{1-\lambda(t)}
\end{array}\right];\,\, E_{1}^{A}=\left[\begin{array}{cc}
0 & \sqrt{\lambda(t)}\\
0 & 0
\end{array}\right],\label{eq:Krauss-amplitude-damping}
\end{equation}
where  \begin{align}
\lambda(t) =1- e^{- g t} \left( \frac{g }{l} \sinh \left[\frac{l t}{2}\right] + \cosh \left[\frac{l t}{2}\right] \right)^2,
\label{eq:lambda}
\end{align}
with $ l = \sqrt{g^2 - 2 \gamma g} $. Here, $g$ is the spectral band width of the noise and $\gamma$ is the system-environment coupling strength. One readily sees that the system exhibits Markovian and non-Markovian evolution when $2 \gamma \ll g$ and $ 2 \gamma \gg g$, respectively \cite{breuer2016colloquium}. 

The above noise may be simulated in an all-optical setup \cite{passos2019non-Markovianity,salles2008experimental,fanchini2014non-Markovianity} by associating the qubit to polarization degrees and the reservoir to the path degrees. With a suitable mapping of the parameters of JC model to the parameters of the optical setup, one may obtain Markovian and non-Markovian effects experimentally. Interestingly, similar to \cite{passos2019non-Markovianity}, the authors of \cite{yugra2020coherence} propose an optical simulation of Markovian and non-Markovian AD. However, we consider the former approach for our case in this paper.



\subsection{Case 1: Only travel qubit subjected to NMAD \label{sec:3a}}
When the photon  returns  back  to Bob,  the state of the system ${hty}$ for either encoding `$j$' can be shown to
have support  of dimensionality 4,  spanned by the  states $\ket{000},
\ket{010},  \ket{100}$ and  $\ket{011}$,  with the  state  of the  $x$
particle being $\ket{2}$, as in  the noiseless attack case. 

After receiving the returned noisy travel qubit, Bob further subjects it to the damping noise, described by Eq. (\ref{eq:Krauss-amplitude-damping}). Accordingly, the final
states with Bob for Alice's encodings $j=0$ and $j=1$ are:
\begin{eqnarray}
\rho^{j=0} &:= \frac{1}{2} \left(
\begin{array}{cccc}
\lambda  & 0 & 0 & 0 \\
0 & 1-\lambda  & \sqrt{1-\lambda } & 0 \\
0 & \sqrt{1-\lambda } & 1 & 0 \\
0 & 0 & 0 & 0 \\
\end{array}
\right);~~
\rho^{j=1} &:= \frac{1}{2} \left(
\begin{array}{cccc}
\lambda  & 0 & 0 & 0 \\
0 & 0 & 0 & 0 \\
0 & 0 & 1 & \sqrt{1-\lambda } \\
0 & 0 & \sqrt{1-\lambda } & 1-\lambda  \\
\end{array}
\right).
\label{eq:AD}
\end{eqnarray}
From Eq. (\ref{eq:AD}), we obtain the  following joint probabilities
$P_{AEB}$:
\begin{align}
&P_{000} = \frac{(\sqrt{1-\lambda}+1)^2}{8} ;\quad P_{001} =  \frac{(\sqrt{1-\lambda}-1)^2}{8}, \nonumber \\
&P_{002} = P_{003}=P_{102} =P_{103} = \frac{ \lambda}{8}; \quad
P_{100} = P_{101} =\frac{1}{8}, \nonumber \\
& P_{110} = P_{111} = \frac{(1-\lambda)}{8},
\label{eq:ADjp2}
\end{align}		
with all  other joint probability  terms vanishing.  Note that  in the
presence of amplitude damping noise, Bob  will also obtain outcomes $|\phi^\pm\rangle
=   \frac{1}{\sqrt{2}}(\ket{00} \pm \ket{11})$    in   his    Bell   state
measurement,  which corresponds  to  the  outcome symbols  2  and 3  in
Eq. (\ref{eq:ADjp2}).

\begin{figure}
	\centering
	\includegraphics[width=0.8\textwidth]{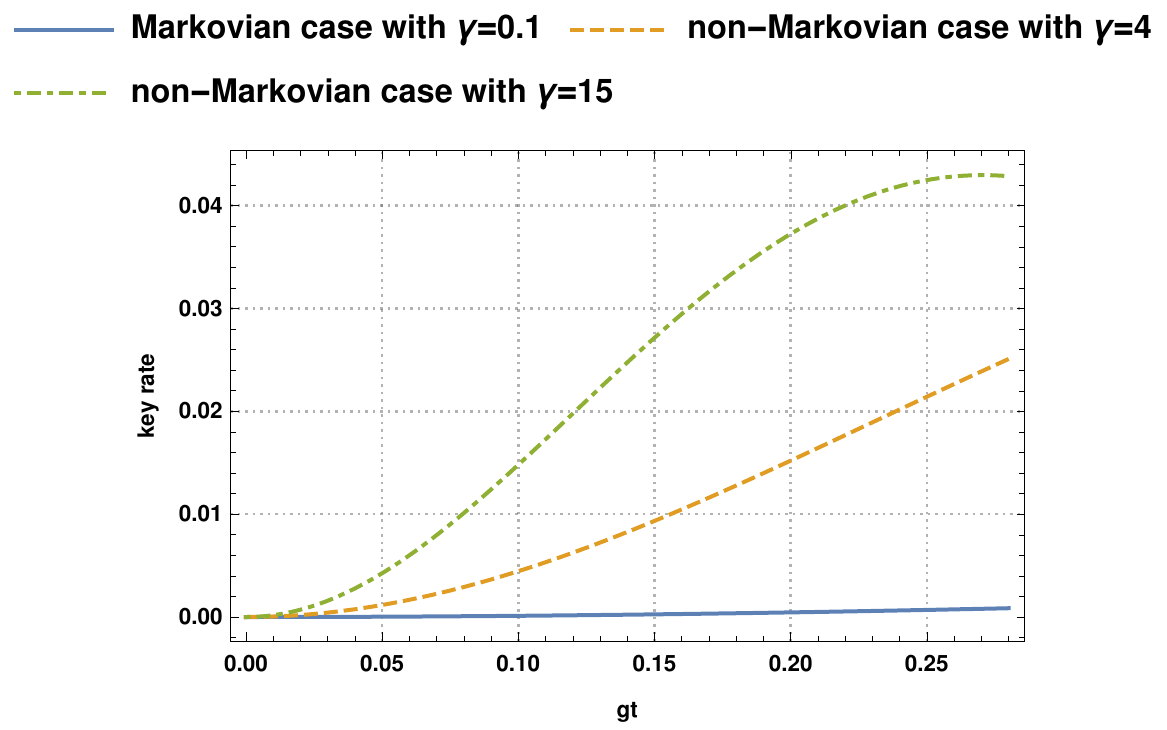}
	\caption{(Color online).  Plot of secure key rate as a function of the dimensionless time $gt$, for the Case 1, where the travel qubit alone is subject to NMAD. Here $\gamma$ is the coupling strength and $g :=1$ in all the cases. In the considered time range, non-Markovian noise provides improvement in the key rate as seen for the cases of $\gamma = 4$ (dashed, orange curve) and $\gamma = 15$ (dot-dashed, green curve), as opposed to the Markovian case with $\gamma = 0.1$ (bold, blue curve).}
	\label{fig:keyrateplot2}
\end{figure}

The probabilities Eq. (\ref{eq:ADjp2}) imply the mutual
information between Alice  and Bob is
\begin{align}
I(A:B) &= - \frac{1}{8}\left( -2\lambda  +\left(\lambda -2 \left(\sqrt{1-\lambda }+1\right)\right) \log \left(\frac{-\lambda +2 \sqrt{1-\lambda }+2}{-\lambda +\sqrt{1-\lambda }+2}\right) \right. \nonumber \\& \left. +(\lambda -2) \log \left(\frac{\lambda -2}{\lambda -\sqrt{1-\lambda }-2}\right)+(\lambda -2) \log \left(\frac{\lambda -2}{\lambda +\sqrt{1-\lambda }-2}\right)\right. \nonumber \\& \left. +\lambda  \log \left(\frac{\lambda +2 \sqrt{1-\lambda }-2}{\lambda +\sqrt{1-\lambda }-2}\right) +2 \left(\sqrt{1-\lambda }-1\right) \log \left(\frac{\lambda +2 \sqrt{1-\lambda }-2}{\lambda +\sqrt{1-\lambda }-2}\right) \right),
\label{eq:ADIAB}
\end{align}
while that between Alice  and Eve:
\begin{align}
I(A:E) &= \frac{2 \log \left(\frac{2}{\lambda +3}\right)+(\lambda +1) \log \left(\frac{\lambda +1}{\lambda +3}\right)+\log (16)}{\log (16)}. 
\end{align}
A plot of the key rate $\kappa  \equiv I_{AB}  - I_{AE}$ {\it w.r.t} (dimensionless) time is given in Figure \ref{fig:keyrateplot2}.

\bla 

\subsection{Case 2: Both the qubits are subject to NMAD \label{sec:3b}}
After receiving the returned noisy travel qubit, Bob subjects both qubits individually to NMAD, described by Eq. (\ref{eq:Krauss-amplitude-damping}). Accordingly, the final
states with Bob for the Alice's encodings $j=0$ and $j=1$ are: 
\begin{eqnarray}
\rho^{(j=0)}_{hty} &= \frac{1}{2} \left(
\begin{array}{cccc}
2 \lambda  & 0 & 0 & 0 \\
0 & 1-\lambda  & 1-\lambda  & 0 \\
0 & 1-\lambda  & 1-\lambda  & 0 \\
0 & 0 & 0 & 0 \\
\end{array}
\right) \,; \quad
\rho^{(j=1)}_{hty} &= \frac{1}{2} \left(
\begin{array}{cccc}
0  & 0 & 0 & 0 \\
0 & 1-\lambda  & 1-\lambda  & 0 \\
0 & 1-\lambda  & 1-\lambda  & 0 \\
0 & 0 & 0 & 2 \lambda \\
\end{array}
\right).
\label{eq:AD1}
\end{eqnarray}
From Eq.  (\ref{eq:AD1}), we obtain the  following joint probabilities
$P_{AEB}$, as follows: \begin{align}
P_{000} &= \frac{1-\lambda}{2}, \nonumber \\
P_{002} &= P_{003}=P_{102} =P_{103} = \frac{ \lambda}{4}, \nonumber\\
P_{100} &= P_{101} =\frac{1- \lambda}{8}, \nonumber\\
P_{110} &= P_{111} = \frac{1-\lambda}{8},
\label{eq:nmadjp}
\end{align}		
with all  other joint probability  terms vanishing.  

\begin{figure}
	\centering
	\includegraphics[width=0.8\textwidth]{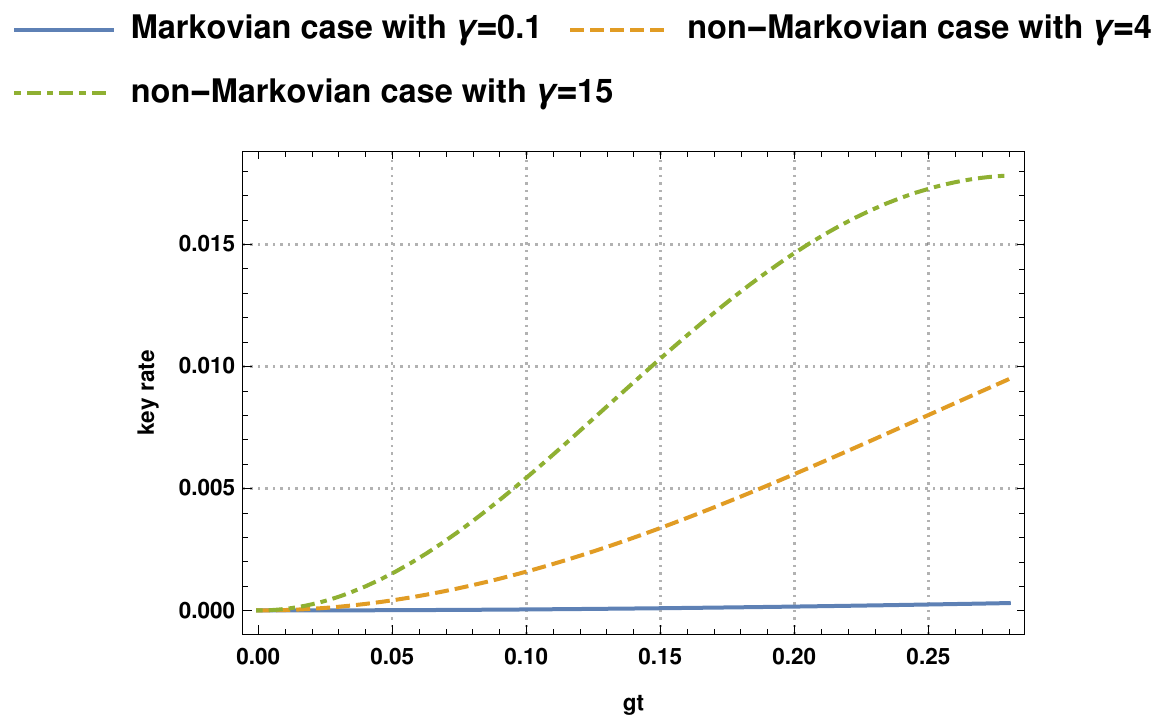}
	\caption{(Color online) Plot of secure key rate with respect to the dimensionless time $gt$, for the Case 2, where the both travel and home qubits are subject to NMAD noise. Here $\gamma$ is the coupling strength and $g :=1$ in all the cases. In the considered time range, non-Markovian noise provides improvement in the key rate as seen for the cases of $\gamma = 4$ (dashed, orange curve) and $\gamma = 15$ (dot-dashed, green curve), as opposed to the Markovian case with $\gamma = 0.1$ (bold, blue curve).}
	\label{fig:keyrateplot}
\end{figure}

From the above probabilities  $P_{AEB}$, one derives  the mutual
information between Alice  and Bob and that between Alice  and Eve, to
be
\begin{align}
I(A:B) &= \frac{3 }{4}(1-\lambda) \log \left(\frac{4}{3}\right) = 0.31(1-\lambda), \nonumber \\
\label{eq:ADIAB}
I(A:E) &= 1+ \frac{1}{2} \log \left(\frac{2}{\lambda +3}\right)+\frac{1}{4} (\lambda +1) \log \left(\frac{\lambda +1}{\lambda +3}\right).
\end{align}
The key rate $\kappa  \equiv I_{AB}  - I_{AE}$  is shown in the Figure (\ref{fig:keyrateplot}).

For both the cases above, from Eqs. (\ref{eq:AD}) and (\ref{eq:AD1}), one can calculate the Holevo bound for Alice-Bob by tracing out Eve's systems $x,y$. It is found that mutual information between Alice and Bob $I(A:B)$ is always lesser than the Alice-Bob Holevo bound suggesting that Bob's measurement strategy is sub-optimal. However, the Holevo bound between Eve's states for Alice's encoding $j\in \{0,1\}$ equals $I(A:E)$, with or without added noise, suggests that Eve's attack strategy in indeed optimal.

\section{On the classical simulation of the quantum advantage \label{sec:proof}}

In Ref. \cite{renner2005information} it was shown that adding classical noise to measurement data by a trusted party  can improve information security. In contrast, here we show that this is not possible for the cases of quantum advantage reported in Sections \ref{sec:3a} and \ref{sec:3b}. That is adding classical noise locally on the part of Bob or even Alice cannot reproduce the benefit of adding the quantum noise. This non-simulability of the quantum advantage may be attributed to the fact that in the regime where the quantum noise is beneficial, it leaves the Bell pair entangled, and thus, the resulting joint probability statistics cannot be captured by local classical noise. 

Consider that Alice and Bob try to locally (i.e., with no communication whatsoever) reproduce $P^{AEB}_{002}, P^{AEB}_{012}$ and $P^\text{AEB}_{112}$ of joint probabilities (\ref{eq:nmadjp}) from the noiseless data (\ref{eq:noiselessjp}). Let $a_{jk}$ define the probability with which Alice uses a pseudo-random number generator (PRNG) to make a transition from a bit value of $A$ in the noiseless data (\ref{eq:noiselessjp})  to a bit value of $A'$ in the noisy data (\ref{eq:nmadjp}), where $A'$ is the bit value locally reproduced by Alice. Similarly, we define the probability $b_{jk}$ for Bob's local transitions using a PRNG to produce a bit value of $B'$. Consider the case of reproducing the following joint probabilities from Eqs. (\ref{eq:noiselessjp}) and (\ref{eq:nmadjp}):
\begin{align}
P^{A'EB'}_{0'12'} &= P^{AEB}_{110}a_{10}b_{02} + P^{AEB}_{111}a_{10}b_{12} = 0 \nonumber \\
&= \frac{a_{10}}{8}(b_{02}+b_{12}) = 0 ,
\label{eq:pab1}
\end{align}
\begin{align}
P^{A'EB'}_{1'12'} &= P^{AEB}_{110}a_{11}.b_{02} + P^{AEB}_{111}a_{11}b_{12} = 0  \nonumber \\
&=\frac{a_{11}}{8}(b_{02}+b_{12}) = 0,
\label{eq:pab2}
\end{align}
and \begin{align}
P^{A'EB'}_{0'02'} &= P^{AEB}_{000}a_{00}b_{02} + P^{AEB}_{100}a_{10}b_{02}+ P^{AEB}_{101}a_{10}b_{12}  \nonumber \\
&= \frac{a_{00}b_{02}}{2} +\frac{a_{10}}{8}(b_{02}+b_{12}) = \frac{\lambda}{4}.
\label{eq:pab3}
\end{align}
From Eq. (\ref{eq:pab1}), it is implied that either $a_{10} = 0$ or $b_{02}+b_{12} =0$ or both are zero. Note that since $\sum_k a_{jk} = 1$, $a_{11} = 1$.  This implies that if $a_{10} = 0$ then, from Eq. (\ref{eq:pab2}), necessarily $b_{02}+b_{12} =0$.

Now, from  Eqs. (\ref{eq:pab1}) and (\ref{eq:pab3}), 
\begin{align}
a_{00}.b_{02} = \frac{\lambda}{2}
\label{eq:pab4}
\end{align}
which implies that $a_{00} \ne  0$ and $b_{02} \ne 0$. Hence we arrive at a contradiction that  $b_{02}+b_{12} \ne 0$.

Now consider that $a_{10} \ne 0$ and $a_{11} \ne 0$. Then from Eq. (\ref{eq:pab1}) and (\ref{eq:pab2}), necessarily $b_{02}+b_{12} =0$. Again from Eq. (\ref{eq:pab3}) and (\ref{eq:pab4}), observe that $b_{02} > 0$. Hence, we arrive at a contradiction again. It follows that Alice and Bob can not unilaterally simulate the quantum advantage due to the NMAD channel by adding uncorrelated local classical noise to their measurement data.

\section{Effect of temperature}
A generalized amplitude damping (GAD) channel models the effect of temperature of the bath along with damping on the qubit state. As in our previous work \cite{sharma2018decoherence}, here we find that unital noise favors Eve in this scenario. We show below that an increase in temperature leads to an increase in the unitality of the channel, and correspondingly to a greater disadvantage for Alice and Bob. One way to understand this effect is as follows. A qubit channel $\mathcal{E}$ is unital if $\mathcal{E}[I] = I$, where $I =\left(
\begin{array}{cc}
1 & 0 \\
0 & 1 \\
\end{array}
\right) $. Now, one may compute
\begin{align}
\rho_{\rm id} = \mathcal{E}^{{\small \rm GAD}}[I] = \left(
\begin{array}{cc}
1 -2 p \lambda +\lambda& 0 \\
0 & (2 p-1) \lambda +1 \\
\end{array}
\right),
\end{align}
where $p \in \{0,\frac{1}{2}\}$.
The action of a GAD channel $ \mathcal{E}^{{\small \rm GAD}}$ on a qubit is given by the quantum operation representation $\mathcal{E}[\rho] = \sum_k A_k \rho A_k^\dagger$, where the $A_k$ are the Kraus operator, which for GAD take the form
\begin{align}
A_1 &= \sqrt{1-p} \left(
\begin{array}{cc}
1 & 0 \\
0 & \sqrt{1-\lambda } \\
\end{array}
\right); \quad
&A_2 = \sqrt{1-p} \left(
\begin{array}{cc}
0 & \sqrt{\lambda } \\
0 & 0 \\
\end{array}
\right) ; \nonumber \\
A_3 &= \sqrt{p} \left(
\begin{array}{cc}
0 & 0 \\
\sqrt{\lambda } & 0 \\
\end{array}
\right); 
\quad
&A_4 =\sqrt{p} \left(
\begin{array}{cc}
\sqrt{1-\lambda } & 0 \\
0 & 1 \\
\end{array}
\right),
\end{align}
where the noise mixing $p \in \{0,\frac{1}{2}\}$ and the damping parameter $\lambda \in \{0,1\}$.

Note that the trace distance (TD) between $\rho_{\rm id}$ and $I$ evaluates to $(2p-1)\lambda$, so that as $p \rightarrow \frac{1}{2}$, the TD $\rightarrow 0$, i.e., $\rho_{\rm id} \rightarrow I$. Therefore increasing temperature enhances the unital part of the noise.

\section{Discussions and Conclusions \label{sec:conclu}}

We consider a QKD based on the Ping-Pong communication protocol, with a non-unital non-Markovian noise deliberately added by the legitimate party before measurement and prior to key distillation. The noise used is the non-Markovian amplitude damping (NMAD). We show that adding this noise improves the security, when Eve uses an optimal individual attack. Conservatively, all the channel noise is attributed to Eve's attack. Within a noise parameter range, non-Markovianity is shown to boost the key rate.  We considered two cases. In one, Bob adds noise only to the travel qubit, whilst in the other, noise it is added to both the travel and home qubits. The former is shown to lead to a higher key rate than the latter in the considered range of time. This provides a cautionary indication that the benefits of non-Markovianity of the noise are conditional and depend on the full context considered. We also studied a non-Markovian generalized amplitude damping (GAD) noise in this context, but in this case we found that temperature tends to diminish the quantum advantage. 

In the matter of local classical non-simulability of the quantum advantage of the considered non-Markovian noise, it is important to stress that the model of classical noise considered in Section \ref{sec:proof} is Markovian, in that at each round the random bit assignment depends only on the measurement outcomes of the current round, and does not require memory of the data from previous rounds. This is a non-trivial assumption, but one that is natural in the current scenario, where the Bell pair used in each round is uncorrelated with any other pair, and furthermore we restrict Eve to attacks on individual qubits. This ensures that the measurement probabilities in each round are independent. Therefore, one expects that classical memory across rounds is not advantageous for the simulation. It is an interesting question whether non-Markovian classical noise can perform better than Markovian classical noise,  if one or both of the above assumptions are relaxed. That is, the protocol may involve Bob's travel qubits being entangled across the rounds and/or Eve launching a joint or collective attack on multiple travel qubits.

Here it may be pointed out that the quantum noise models given by Eqs. (\ref{eq:Krauss-amplitude-damping}) and (\ref{eq:lambda}) are considered non-Markovian despite being applied to individual rounds of the protocol. The reason is that the memory in this context is with respect to an external environment, rather than preceding rounds of the protocol. In particular, quantum non-Markovianity arises when the dynamics of the system-environment correlation makes the system's intermediate map (or, propagator) to deviate from complete positivity. \cite{RHP14}.

\section*{Acknowledgments}

SU and RS acknowledge financial support of the Govt. of India DST-SERB grant EMR/2016/004019. SU also thanks the Admar Mutt Education Foundation (AMEF), Bengaluru, Karnataka, India for partial financial support. SB and RS acknowledge the support from Interdisciplinary Cyber Physical Systems (ICPS) programme of the Department of Science and Technology (DST), India, Grants No.: DST/ICPS/QuST/Theme-1/2019/6 and DST/ICPS/QuST/Theme-1/2019/14, respectively. US thanks Ashutosh Singh and S. Omkar for helpful discussions.

\bibliographystyle{spbasic}

\end{document}